\documentclass[screen,authorversion]{acmart} %
\AtBeginDocument{%
  }

\usepackage{xcolor}
\definecolor{mygreen}{RGB}{0, 150, 0} 
 \usepackage{multirow}
 \usepackage{hyperref}
 \usepackage{soul}
 \sethlcolor{lightgreen}
 \usepackage{soul} %
\usepackage{xcolor} %
\usepackage{float}

\definecolor{lightgreen}{RGB}{144, 238, 144}

\sethlcolor{lightgreen}

\copyrightyear{2025}
\acmYear{2025}
\setcopyright{rightsretained}
\acmConference[LAK 2025]{LAK25: The 15th International Learning Analytics and Knowledge Conference}{March 03--07, 2025}{Dublin, Ireland}
\acmBooktitle{LAK25: The 15th International Learning Analytics and Knowledge Conference (LAK 2025), March 03--07, 2025, Dublin, Ireland}
\acmDOI{10.1145/3706468.3706531}
\acmISBN{979-8-4007-0701-8/25/03}

\begin{document}

\title{Do Tutors Learn from Equity Training and Can Generative AI Assess It?}

\author{Danielle R. Thomas}
\email{drthomas@cmu.edu}
\affiliation{
  \institution{Carnegie Mellon University}
  \city{Pittsburgh, PA}
  \country{USA}
}

\author{Conrad Borchers}
\email{cborcher@cs.cmu.edu}
\affiliation{%
  \institution{Carnegie Mellon University}
  \city{Pittsburgh, PA}
  \country{USA}
}

\author{Sanjit Kakarla}
\email{sanjit.kakarla@gmail.com}
\affiliation{%
  \institution{Carnegie Mellon University}
  \city{Pittsburgh, PA}
  \country{USA}
}

\author{Jionghao Lin}
\email{jionghao@cmu.edu}
\affiliation{%
  \institution{Carnegie Mellon University}
  \city{Pittsburgh, PA}
  \country{USA}
}

\author{Shambhavi Bhushan}
\email{shambhab@andrew.cmu.edu}
\affiliation{%
  \institution{Carnegie Mellon University}
  \city{Pittsburgh, PA}
  \country{USA}
}

\author{Boyuan Guo}
\email{boyuan@andrew.cmu.edu}
\affiliation{%
  \institution{Carnegie Mellon University}
  \city{Pittsburgh, PA}
  \country{USA}
}

\author{Erin Gatz}
\email{egatz@andrew.cmu.edu}
\affiliation{%
  \institution{Carnegie Mellon University}
  \city{Pittsburgh, PA}
  \country{USA}
  }
  
\author{Kenneth R. Koedinger}
\email{koedinger@cmu.edu}
\affiliation{
  \institution{Carnegie Mellon University}
  \city{Pittsburgh, PA}
  \country{USA}
}

\renewcommand{\shortauthors}{Thomas et al.}

\begin{abstract}
 Equity is a core concern of learning analytics. However, applications that teach and assess equity skills, particularly at scale are lacking, often due to barriers in evaluating language. Advances in generative AI via large language models (LLMs) are being used in a wide range of applications, with this present work assessing its use in the equity domain. We evaluate tutor performance within an online lesson on enhancing tutors' skills when responding to students in potentially inequitable situations. We apply a mixed-method approach to analyze the performance of 81 undergraduate remote tutors. We find marginally significant learning gains with increases in tutors' self-reported confidence in their knowledge in responding to middle school students experiencing possible inequities from pretest to posttest. Both GPT-4o and GPT-4-turbo demonstrate proficiency in assessing tutors ability to \textit{predict} and \textit{explain} the best approach. Balancing performance, efficiency, and cost, we determine that few-shot learning using GPT-4o is the preferred model. This work makes available a dataset of lesson log data, tutor responses, rubrics for human annotation, and generative AI prompts. Future work involves leveling the difficulty among scenarios and enhancing LLM prompts for large-scale grading and assessment.
 \end{abstract}



\maketitle

\section{Introduction}
Equity is a core concern of learning analytics (LA) \cite{khalil2023fairness}, however, applications that teach and assess equity competencies are lacking. Intelligent tutoring systems, a core learning analytics application, have improved learning in STEM domains with well-defined rules for assessment. However, equity competencies are taught and assessed through language which are harder to grade, requiring advanced assessment methods. Advances in large language models (LLMs), which demonstrate text comprehension in various domains \cite{chang2024survey}, could help bring the benefits that make tutoring systems effective (e.g., automated assessment and immediate feedback) to the equity domain. To this end, this work introduces the use of generative AI to evaluate human tutors' open-ended responses involving approaches to equity, a novel and under-researched LA application. In addition, we contribute a dataset of lesson log data, human annotation rubrics, and generative AI prompts to enhance transparency, reproducibility, and collaboration within the LAK community.

\textit{Equality} within education refers to providing all students with the same resources and opportunities, regardless of circumstances and despite any inherent advantages or disadvantages. \textit{Equity}, on the other hand, focuses on tailoring support to meet the individual needs of the student, recognizing that the background and challenges of each student are unique \cite{AECF2015}. While equality treats every student alike, equity adjusts resources to ensure all students have the chance to achieve similar success. Social justice education focuses on raising students' consciousness about inequity in everyday social and educational situations \cite{duncan2009note, hammond2021liberatory}. A tutor telling a student, ``if you work hard enough, you will be successful'' may be great advice but only if the student has access, human support, and opportunities to engage with the learning materials. For example, a student without books at home may find it difficult to read when out of school. Tutors can help secondary students manage inequities in their learning by assisting them with recognizing possible inequities and supporting them in advocating for themselves--but how does a tutor go about providing students equity--based support? How can tutors learn and develop advocacy skills? This present work uses human coding and generative AI to evaluate tutor learning within a lesson to help students manage inequities.

The evaluation of open-ended responses within assessments by human graders is costly and time consuming \cite{black2009developing}. Leveraging large language models to automatically assess tutors' open responses holds promise for transforming small-scale instructional activities into large-scale and personalized training programs \cite{lin2024using, thomas2023tutor}. In this study, we discuss evaluating tutors' open-ended responses using LLMs. Previous research has explored the use of LLMs to assess tutor learning on skills related to social emotional learning, such as giving effective praise \cite{hanimproving} and providing content support such as reacting to students making math errors \cite{kakarla2024using}. Here, we leverage this method and adapt it to equity-focused tutor training. Ultimately, this present work uses generative AI, specifically the large language models GPT-4-turbo and GPT-4o, to develop a method to assess open responses of tutors while participating in scenario-based training. We intend to develop a systematic, scalable approach to provide real-time assessment.

The present work is of great interest to the LAK community through the identification of evidence of learning skills and their assessment. Using lesson log data, we use open-ended responses and multiple-choice selections of the tutor to analyze and determine tutor learning gains. Adding data from self-reported tutor surveys, we determine the construct validity of equity-focused training and the perceptions of tutors about their learning. Expanding the LAK methodological toolbox, we leverage a novel use of generative AI allowing LLMs to assess open-ended responses of teachers in the equity domain. Currently, we know of very little past work using generative AI to assess tutor responses within scenario-based training. In this work, we address the following research questions:

\textbf{RQ1}: Is the scenario-based lesson effective in teaching tutors new skills for responding to students possibly experiencing inequities? 

\textbf{RQ2}: How does tutors' self-reported confidence of their knowledge attending to students experiencing potential inequities change from pretest to posttest, and do tutors feel they can apply what they have learned? 

\textbf{RQ3}: How effective are large language models GPT-4o and GPT-4-turbo in assessing tutors' actions in responding to students managing possible inequity? 

\textbf{RQ4}: How do the large language models GPT-4o and GPT-4-turbo compare in performance, efficiency, and cost?

\section{Related Work}
\subsection{The Role of Scenario-based Learning in Tutor Development} Scenario-based learning integrates educational activities within real-world contexts, promoting rapid development of skills through situational activities \cite{bardach2021power}. This instructional strategy has been successfully applied across multiple disciplines, such as medical education, fostering prospective thinking among high school students, and enhancing the growth of pre-service teachers (e.g., \cite{mclean2016case, alattar2019effectiveness, hursen2017investigating}). Furthermore, scenario-based models, including digital simulations, offer novice teachers and tutors valuable low-risk practice environments to gain situational experience \cite{thompson2019teacher, chine2022development, chine2022scenario}. Authentic scenarios, coming from real-life tutoring situations, are used by tutors to practice ``learning by doing,'' which is an instructional approach that emphasizes active participation and hands-on practice to acquire knowledge and skills through direct experience. The intention is that this learning transfers to real-life tutoring environments. Learning by doing requires the application of skills that model the needs of the real world \cite{koedinger2015learning}, facilitating the transfer of new knowledge to similar experiences that trigger recall \cite{schank2013learning}.

\begin{figure*}[ht]
\includegraphics[scale = 0.5]{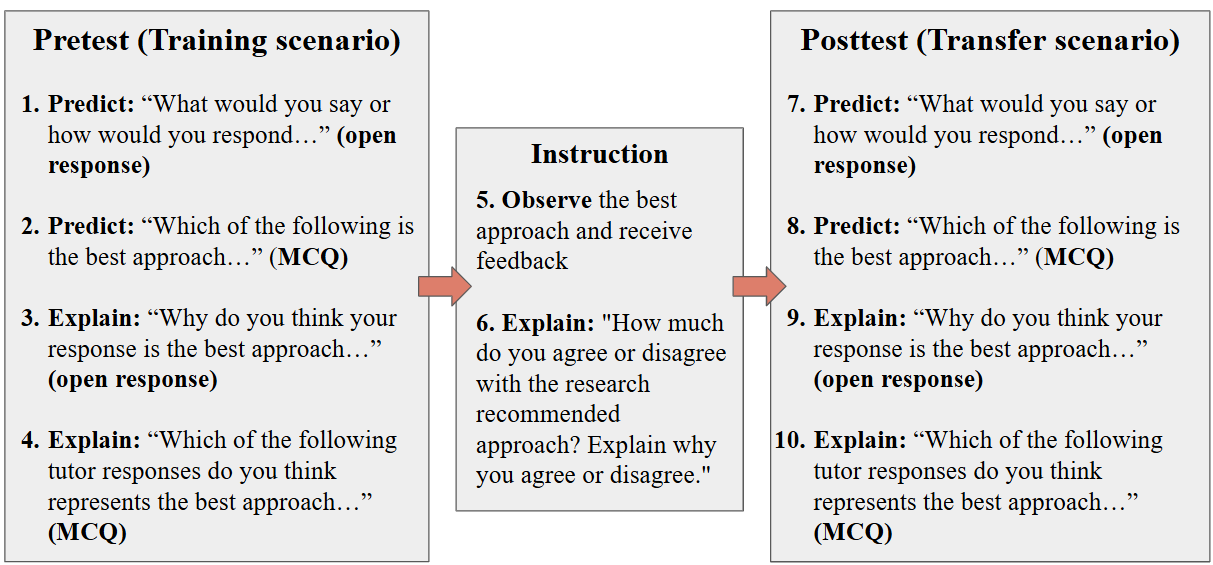}
\caption{The modified predict-observe-explain cycle for the pretest and posttest scenarios. } \label{fig:1_1}
\Description{The modified predict-observe-explain cycle for the pretest and posttest scenarios.}
\end{figure*}

Past studies have shown approximately 20\% learning gain from pretest to posttest in scenario-based lessons covering tutor topics such as giving effective praise to students; reacting when a student makes an error; and determining what students know \cite{thomas2023tutor}. Both training and transfer scenarios (also known as pretest and posttest) follow a modified predict-observe-explain (POE) approach, theoretically connected to Gibbs' Reflective Cycle,  a cyclical instructional model providing structure for learning by doing to individual learning experiences \cite{gibbs1988learning}. In line with Figure \ref{fig:1_1}, tutors respond to a training scenario, or pretest, (that is, a student possibly experiencing an inequity) by asking them to \textit{predict} how to best respond within 1) an open-ended question and 2) a multiple-choice question (MCQ). Then tutors \textit{explain} their prediction via 3) an open-ended question and 4) a MCQ. Tutors then 5) observe the  given research recommendation and receive feedback before they 6) \textit{explain} the reasoning behind what they observed. Finally, the tutors complete the transfer scenario or posttest, following the same pattern of predicting the best responses and explaining their reasoning (7-10). The pretest and posttest each contain a maximum of four points (two MCQs and two open-ended questions), with tutor learning gains determined by subtracting tutor pretest score from the posttest score. This process provides a method for assessing the transfer of learning and, ultimately, the tutor's learning gain \cite{chine2022development, thomas2023tutor}. 
\subsection{Addressing Educational Inequities through Scenario-based Learning}
Ensuring equity remains one of the greatest challenges in education today \cite{AECF2015, gonzalez2017equity}. For example, a common situation that educators and tutors face is a student who cannot complete an assignment or access instructional materials outside of school because they do not have the Internet at home. Among school-age children worldwide, an estimated two-thirds do not have access to the Internet at home \cite{unicef2020many}. Many students do not recognize such inequities when they experience them. A student without the internet at home tasked with completing an assignment that requires broadband access will simply not do the assignment. They may not recognize the lack of access as inequitable and frankly unfair. Educators and tutors can play an important role in promoting equity. Rapid-cycle learning, in the form of short instructional activities for educators and tutors, is gaining traction to exercise applying strategies to advance equitable practices \cite{AECF2015}. Rapid-cycle learning, exemplified by the situational judgment tests presented in this work, offers a method of providing instructional support to educators and tutors in attending to social justice and equity concerns \cite{Shenbanjo2024}. Through brief, scenario-based learning activities, tutors participate in low-risk opportunities to pilot strategies to assist students with diverse experiences and needs, whether or not they are related to systemic inequities. Providing students with high-impact and personalized tutoring is becoming a potential solution to narrow the opportunity gap among underserved students \cite{guryan2023not}. However, there is a shortage of experienced tutors \cite{kraft2021blueprint}, along with limited training in supporting social justice and ensuring equitable learning environments. This work aims to evaluate the learning of the tutor from such activities on how to help middle school students manage potential inequitable situations. 
\subsection{The \textit{Helping Students Manage Inequity} Lesson} One approach for promoting equity-responsive practices among secondary students is instructing students on collaborating with adults and advocating for themselves \cite{gonzalez2017equity}. The lesson draws on previous research to identify the key competencies of effective tutoring, with scenario-based learning activities developed to align with key competencies \cite{thomas2023tutor}. We strive to determine tutor learning gains similar to \cite{thomas2023tutor} while focusing specifically on how tutors respond to students experiencing needs, potentially indicative of inequities. The lesson objectives include: recognizing when a student may be experiencing inequity related to their learning; and applying strategies to help students manage inequities by assisting students to advocate for themselves. 

\begin{figure}[ht]
\includegraphics[width=0.8\textwidth]{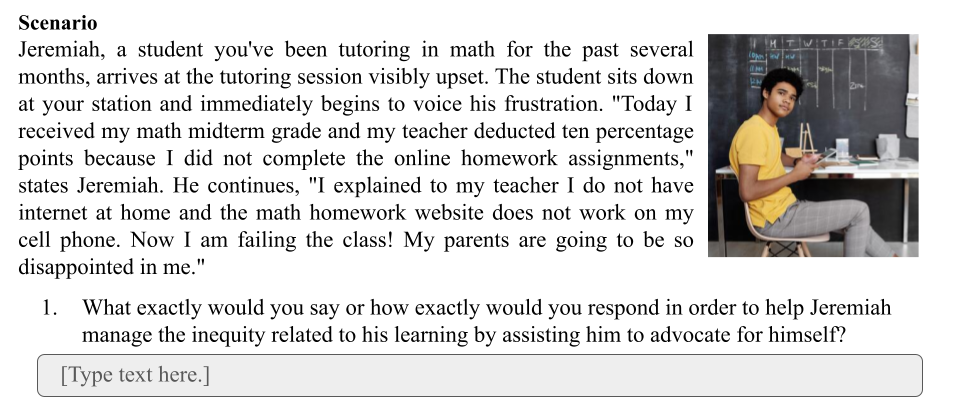}
\caption{The scenario involving student Jeremiah with the open-ended  question prompting a tutor to predict the best approach.} \label{fig:12}
\Description{The scenario involving student Jeremiah with the open-ended  question prompting a tutor to predict the best approach.}
\end{figure}

In one of two scenarios (used interchangeably as a pretest or posttest), student Jeremiah could not complete his homework because he did not have access to the Internet at home. In this specific situation, the tutor does not know if the student did not have the internet for temporary reasons, such as a random system outage, or if the lack of internet access is associated with a systemic disparity affecting the student's ability to succeed, such as socioeconomic status. However, the research-recommended approach of the tutor is to help Jeremiah recognize his need and empower him to advocate for himself by exercising his voice \cite{gonzalez2017equity}. Figure \ref{fig:12} illustrates the scenario involving student Jeremiah showing the open-ended question asking tutors to \textit{predict} the best approach (shown), which is followed by a selected-response question with options for tutors to choose how they would respond (not shown). The tutors are then asked to \textit{ explain} why they chose their choice within an open response and multiple-choice questions.

\begin{figure}[ht]
\includegraphics[width=0.8\textwidth]{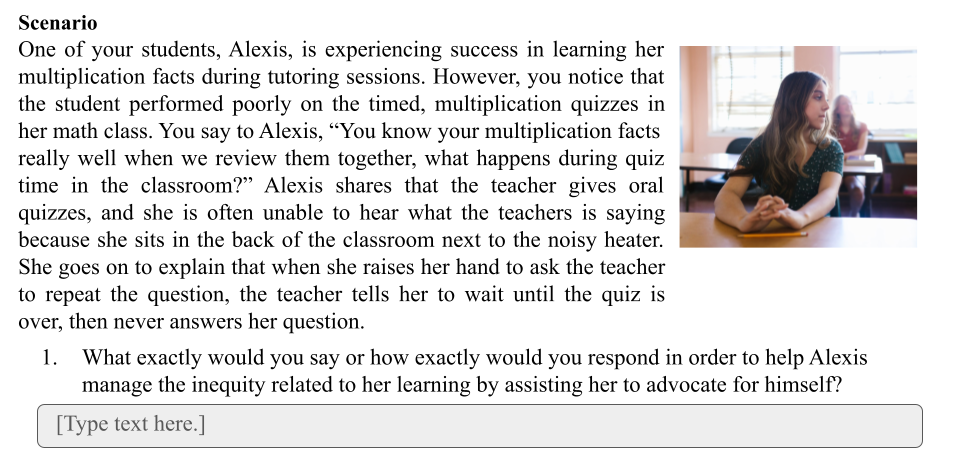}
\caption{The scenario involving student Alexis with the open-ended question prompting a tutor to predict the best approach.} \label{fig:2}
\Description{The scenario involving student Alexis with the open-ended question prompting a tutor to predict the best approach.}
\end{figure}

Similarly, an analogous scenario details the situation of Alexis shown in Figure \ref{fig:2}, who expresses to her tutor that she earned a bad grade on a math assignment because she sits in the back of the classroom and cannot hear. Both the Jeremiah and Alexis scenarios and all questions are available in the \href{https://github.com/CMU-PLUS/LAK2025-Inequity}{Digital Appendix.} Alexis seating issue, such as Jeremiah's lack of Internet access, might seem minor, but could indicate deeper systemic inequities in education, such as inadequate support for diverse learning needs or accessibility challenges. Recognizing and addressing these issues is vital for tutors, as it involves understanding broader educational barriers and developing inclusive strategies to support every student's success. This approach represents a fundamental step towards achieving educational equity, ensuring that all students receive the necessary opportunities and support to thrive, regardless of their background or circumstances.  

\subsection{Benefits of Multiple-choice and Open-Ended Responses}
Multiple-choice questions (MCQs) are a type of closed-response question often used in assessments due to their efficiency and objective grading \cite{butler2018multiple}. However, they come with challenges such as encouraging reliance on test-taking strategies, issues with face validity, and difficulty in generating high-quality distractors or ``incorrect'' options \cite{butler2018multiple}. In contrast, open-ended questions challenge students to develop their own responses, most often through textual language that reduces the influence of guessing, offer improved face validity, and can require higher-order thinking skills more readily \cite{butler2018multiple}. Despite these advantages, grading open-ended responses is resource intensive (e.g., human graders often need to assess learner responses and provide feedback), making them less practical for large-scale use \cite{black2009developing}. Recent advances in automatic short answer grading (ASAG) have shown potential in automating this process, but these models have historically faced challenges when used in different educational settings. Subtle differences between tasks (e.g., tutor training in promoting advocacy and ensuring equity in nuanced social situations) can significantly affect how well the model performs in diverse educational contexts \cite{zhang2022automatic}.

\subsection{Using Generative AI to Assess Open-ended Tutor Responses}Early ASAG approaches for assessing open-ended responses relied on traditional machine learning techniques, such as feature extraction and bag-of-words models, which provided easily interpretable results \cite{zhang2022automatic, madnani2013automated, hou2011automatic}. However, these early models struggled with domain shifts, where subtle differences in assessment tasks significantly impaired their performance \cite{henkel2024can}. More recently, studies have turned to deep learning models for automated assessment of open-ended responses \cite{condor2021automatic, zhang2016deep, tan2023automatic}. For example, \cite{condor2021automatic} explored ASAG using Sentence-BERT (SBERT) to measure textual similarity within learner response grading. Although SBERT showed promise for generalization, it encountered difficulties with unseen questions, revealing the need for models capable of deeper contextual understanding and adaptability, particularly in more complex domains such as equity training.

Recent advances in generative AI, particularly large language models (LLMs) such as GPT-4, have revolutionized ASAG by enabling models trained on extensive datasets to interpret and assess nuanced textual responses \cite{chang2024survey}, with applications in tutor training \cite{hanimproving, lin2024can, kakarla2024using}. LLMs offer flexibility, as they can be fine-tuned for specific educational tasks or used in their pre-trained form. Studies show that, with effective prompting, LLMs can capture the subtleties in learner responses, improving grading efficiency and scalability \cite{henkel2024can}. Nevertheless, challenges persist, as LLMs operate as "black-box" models, making their outputs difficult to interpret \cite{chang2024survey}. Furthermore, LLMs lack consistent knowledge of the pedagogy and content and are prone to hallucinations, generating confidently inaccurate or nonsensical information \cite{zhang2022automatic}. There remains an urgent need to explore how LLMs can be trained to assess equity-focused tutor responses, such as recognizing when tutors advocate for underserved students or support them in addressing inequities. These advancements could significantly improve tutor training to promote equitable education.

\section{Method}
\subsection{Tutor Participants, Lesson Delivery, and Construct Validity}
There were 81 college-student participants who completed the lesson, employed as paid tutors for a remote tutoring organization supporting middle-school students. While the demographics of the tutors were undisclosed, they exhibited cultural and racial diversity. Tutors' self-reported tutor experience levels were assessed using a 5-point Likert scale with 1 indicating little to no experience (novice) and 5 indicating an expert tutor. On average, the tutors reported an experience level of 3.2 (\textit{SD} = 1.22). Furthermore, measures of tutor self-perceptions of their confidence in the topic were surveyed before pretest and after posttest using a 5-point Likert scale ranging from 1 (\textit{not at all confident}) to 5 (\textit{extremely confident}). After the posttest, tutors were surveyed regarding the application of the lesson content using a 5-point Likert scale (1- \textit{strongly disagree}; 5- \textit{strongly agree}): \textit{I am confident I can apply what I learned}. The lesson was crafted by a university research team specializing in learning science, in collaboration with an equity-focused consulting firm, enhancing construct validity. The lesson was delivered through an online tutoring platform and is consistent with the research-shown competencies of effective tutoring within the area of \textit{Advocacy} \cite{thomas2023tutor, PLUS2024}. We prioritized maintaining the privacy and confidentiality of tutors, adhering to all Institutional Review Board (IRB) requirements.

\subsection{Human Open-ended Response Coding}
Open-ended questions were coded, with ``correct'' tutor responses designated as ``1'' and ``incorrect'' as ``0.'' Two experienced researchers coded participant responses to assess interrater reliability. The responses were deemed correct if the tutors demonstrated understanding of the lesson objectives and the research-recommended approach. Tables \ref{tab:my-table1} and \ref{tab:my-table2} present responses sourced from the learners with a rationale for the coding to predict and explain the responses, respectively. For \textit{predicting} the best response, or correct response (score = 1), tutors should apply the strategy of assisting the middle or high school student in encouraging them to advocate for themselves by talking directly with their teacher or person in charge. In contrast, incorrect responses (score = 0) involved the tutor suggesting that the student find a way to solve the problem on their own or made suggestions that do not directly involve the student advocating for themselves in attending to their needs. Incorrect responses also include responses where the tutor merely indicates to a student to speak with a teacher without encouraging active advocacy. For \textit{explaining} their chosen approach, tutor responses must demonstrate that the tutor recognizes that the student needs support in advocating for themselves and encouraging the student to act.

\begin{table}[ht]
\caption{Learner-sourced responses for \textit{predicting} the best approach with rationale. Utterances aligned with correct, or ``desired,'' rationale are highlighted in green.}
\label{tab:my-table1}

\resizebox{0.8\textwidth}{!}{%
\renewcommand{\arraystretch}{1.6} %
\fontsize{12}{12}\selectfont     %
\begin{tabular}{p{7.0cm}p{10.0cm}} %

\hline
\textbf{Tutor Response} & \textbf{Coding Rationale} \\ \hline
\textit{Let's work together to come up with a solution and find a way to \hl{address this issue with your teacher}, so you have a \hl{fair opportunity to succeed.}} & Correct (1): This response assists the student with recognizing possible inequity related to their learning and helps the student in advocating for themselves. \\\hline
\textit{I would provide him with other possible ways to access the internet that helps him do the homework equally. For example, he could do the homework at school.} &  Incorrect (0): This response recommends to the student alternative approaches to solve the problem but does not directly promote student advocacy. \\\hline
\textit{I am very sorry to hear this. I understand that you are upset and I am upset about it too. However, unfair things happen in life, so we should find ways to resolve them in our own situations.} & Incorrect (0): Although this response validates the student's feelings, it does not help the student recognize possible inequity nor help the student advocate for themselves. \\ \hline
\end{tabular}
}
\end{table}

\begin{table}[ht]
\caption{Learner-sourced responses for \textit{explaining} their chosen approach with rationale. Utterances aligned with correct, or ``desired,'' rationale are highlighted in green.}
\label{tab:my-table2}

\resizebox{0.8\textwidth}{!}{%
\renewcommand{\arraystretch}{1.6} %
\fontsize{12}{12}\selectfont     %
\begin{tabular}{p{7.0cm}p{10.0cm}} %

\hline
\textbf{Tutor Response} & \textbf{Coding Rationale} \\ \hline
\textit{It will allow Alexis to \hl{go over the problem herself} and \hl{allows her to feel empowered.}} & Correct (1): This response indicates the tutor recognizes the importance of the student advocating for themselves. \\\hline
\textit{It encourages him to \hl{make a plan to try and fix the problem, rather than just encouraging him to "work harder"} when hard work simply will not resolve the issue.} &  Correct (1): This response indicates the tutor recognizes the importance of the student advocating for themselves.  \\\hline
\textit{Actively trying to come up with a solution is more important than simply expressing sorrow.} &Incorrect (0): This response does not indicate the tutor recognizes the importance of student advocacy but merely focuses on problem solving with student. \\ \hline
\end{tabular}
}
\end{table}

\subsection{Inter-rater Reliability Among Human Graders}
Human coders assessed the responses of all 81 tutors for the \textit{predict} and \textit{explain} open-ended responses. The results indicated a relatively high agreement in inter-rater reliability between the coders using the binary coding system (i.e., 0, 1). For responses requiring tutors to \textit{predict} the best course of action, there was 89\% agreement and a Cohen's $\kappa$ of 0.75. For responses asking tutors to \textit{explain} their rationale, there was an 87\% agreement rate and a Cohen's $\kappa$ of 0.73. Both reflect substantial agreement, supporting the reliability of the coding process.

\subsection{Determining Tutor Learning Gains}
We employed a mixed-effects ANOVA to examine the impact of lesson scenarios on tutor performance. The \textit{scenarios} (i.e., Jeremiah or Alexis) served as the between-subjects factor, while time, specifically \textit{pretest} and \textit{posttest}, served as the within-subjects factor. Treating \textit{ scenario} as a fixed effect helps to determine if there is an imbalance in difficulty between the two scenarios, while considering the test \textit{ time} as a random effect accounts for variation within the subjects. The ANOVA was run on data of students who completed all training and test items, resulting in a reduced sample of 81 students. A split-half reliability analysis adjusting for a time factor between pre and post-test was used to determine the reliability of the employed test battery. Specifically, we correlated person parameters of Rasch models across all possible splits of items. This resulted in an average reliability of 0.489 across all assessment items, which is acceptable for a test including only eight assessment items, although lower than typical test batteries which feature more items \cite{kim2010estimation}.

\subsection{Prompt Engineering to Evaluate Tutor Responses } 
Drawing from recent advancements in prompt engineering, particularly within the domain of tutor training \cite{lin2024using, thomas2024learning}, we developed prompts to leverage GPT-4-turbo and GPT-4o for evaluating the correctness of open-ended tutor responses. Model temperature was set to 0 to ensure the model operated deterministically, selecting the most likely next word (or token) based on the input, reducing variability. This approach leads to more predictable and generally more cautious responses. The output was limited to 300 tokens to avoid excessive verbosity. Using zero-shot and few-shot prompting approaches, we provided the models with examples to guide their assessment of the predict and explain responses. The creation of these prompts followed an iterative process, with several rounds of adjustments informed by feedback from initial model outputs. Table \ref{tab:promtp_praise1} and \ref{tab:promtp_praise2} demonstrate the specific few-shot learning prompts designed for these tasks. 

We employed several prompt engineering strategies to enhance model performance, including chain-of-thought prompting \cite{wei2022chain} to encourage step-by-step reasoning by prompting the model to provide the rationale. We used few-shot prompting (Brown et al., 2020), by supplying the model with relevant examples of correct and incorrect tutor responses to guide its output. Last, we used contextual priming to help the model understand the task within a specific context (i.e., \textit{``You are a tutor evaluator…''}). Together, these techniques enabled the models to assess tutor responses with greater accuracy and consistency, aligning outputs more closely with human evaluations and enhancing overall model performance.

\begin{table*}[ht]
\caption{Complete prompt for GPT-4o (few shot) used for the task of assessing tutors in \textit{predicting} the best approach.}
\label{tab:promtp_praise1}
\resizebox{\textwidth}{!}{%
\renewcommand{\arraystretch}{1.5} %
\fontsize{11}{13}\selectfont %
\begin{tabular}{p{18cm}} %
\hline
\small
\begin{tabular}[c]{@{}p{18cm}@{}}  \texttt{SCORING\_PROMPT\_START} = """\\ Please assess a tutor's response in a tutor training scenario involving a middle school student struggling to understand a math problem.

-if the tutor's response helps the student identify that they have a need and provides the student support on how to remedy, by dealing with the issue and promoting student advocacy, or the tutor assists the student in helping them talk with their teacher without providing an alternative solution or work around, score with a 1. Examples of responses scoring a 1 are: "Jeremiah can advocate for improved technology access"; and "Jeremiah, you are doing a good job trying to communicate with your teacher. As she doesn't care, I can provide you several ways to let her not treating you unfairly."

-if the tutor's response does not demonstrate that the tutor understands that the student needs support in advocating for themselves, score with a 0. Sample responses scoring a 0 include: ``That would be the right approach because I am helping the student solve the problem rationally"; "This can let him feel supported and at the same time, gives him a solution of this problem"; "Because it teaches Alexis a possible way to avoid similar problems in the future; and "Discussing the plan with the student will provide them how to stand up for the inequity."

Response Start {-}{-}{-}\\ """\\ \texttt{FORMAT\_PROMPT} = \\ "{-}{-}{-} Response End. Given the earlier transcript, please return a JSON string following the format, \{\textbackslash{}"Rationale\textbackslash{}": \textbackslash{}"your reasoning here\textbackslash{}", \textbackslash{}"Score\textbackslash{}":0/1\}."\end{tabular} \\ \hline 
\end{tabular}
}
\end{table*}

\begin{table*}[ht]
\caption{Complete prompt for GPT-4o (few shot) used for the task of assessing tutors in \textit{explaining} the best approach.}
\label{tab:promtp_praise2}
\resizebox{\textwidth}{!}{%
\renewcommand{\arraystretch}{1.5} %
\fontsize{11}{13}\selectfont %
\begin{tabular}{p{18cm}} %
\hline
\small
\begin{tabular}[c]{@{}p{18cm}@{}}  \texttt{SCORING\_PROMPT\_START} = """\\ Please assess a tutor's response in a tutor training scenario involving a middle school student struggling to understand a math problem.

-if the tutor's response demonstrates that they recognize that the student needs support in advocating for themselves and encourages the student to act, score with a 1. Sample responses scoring a 1 include: ``It practices critical hope. It encourages her to advocate for herself by speaking to the teacher about it, and I offer my own help to assist her''; ``It prepares Alexis to solve the problem herself by practicing in a low stress environment. It would make her more confident to talk to her teacher and do better in class.'' 

-if the tutor's response does not demonstrate that the tutor understands that the student needs support in advocating for themselves, score with a 0. Sample responses scoring a 0 include: ``That would be the right approach because I am helping the student solve the problem rationally"; "This can let him feel supported and at the same time, gives him a solution of this problem"; "Because it teaches Alexis a possible way to avoid similar problems in the future; and "Discussing the plan with the student will provide them how to stand up for the inequity."

Response Start {-}{-}{-}\\ """\\ \texttt{FORMAT\_PROMPT} = \\ "{-}{-}{-} Response End. Given the earlier transcript, please return a JSON string following the format, \{\textbackslash{}"Rationale\textbackslash{}": \textbackslash{}"your reasoning here\textbackslash{}", \textbackslash{}"Score\textbackslash{}":0/1\}."\end{tabular} \\ \hline 
\end{tabular}
}
\end{table*}

\section{Results}
\subsection{\textbf{RQ1: Is the scenario-based lesson effective in teaching tutors new skills for responding to students possibly experiencing inequities?}} 
\begin{figure}[ht]
\includegraphics[width=0.8\textwidth]{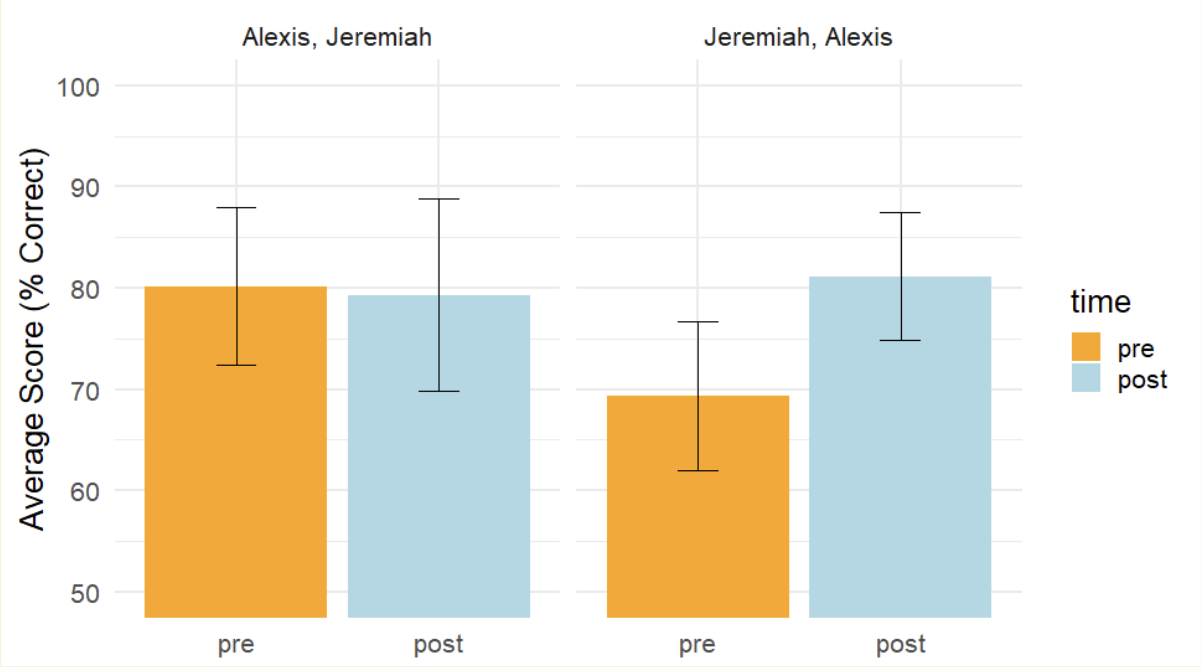}
\caption{Mean pretest and posttest scores between scenario order conditions and measurement points.} \label{fig:135}
\Description{Mean pretest and posttest scores between scenario order conditions and measurement points.}
\end{figure}

The analysis revealed a marginally significant main effect of \textit{time} on tutor performance, $F(1, 79) = 3.20$, $p = .078$, indicating an overall improvement in tutors' performance from pretest to posttest. This suggests a general learning effect or improved skills. This main effect was qualified by a significant interaction between \textit{scenario} and \textit{time}, $F(1, 79) = 4.31$, $p = .041$. This significant interaction implies that the degree of improvement from pretest to posttest varied between the two scenarios. There was no statistically significant main effect of the specific \textit{scenario} on tutor performance, $F(1, 79) = 0.54$, $p = .465$, indicating that the difficulty level between the Jeremiah and Alexis scenarios did not differ significantly in terms of overall tutor performance. \textit{Post-hoc contrasts} were conducted to examine differences between pretest and posttest scores across the two scenario orders based on marginal means. Shining light on the significant interaction, learning gains in the \textit{Alexis:Jeremiah} scenario order were not statistically significant, $M = -0.01$, $SE = 0.05$, $t(79) = -0.02$, $p = .846$. However, learning gains were significant in the \textit{Jeremiah:Alexis} scenario order, $M = 0.12$, $SE = 0.04$, $t(79) = 3.07$, $p = .001$. The differences in assessment scores between the two scenario order conditions are visualized in Figure \ref{fig:135}.

\subsection{\textbf{RQ2: How does tutors' self-reported confidence of their knowledge attending to students experiencing potential inequities change from pretest to posttest, and do tutors feel they can apply what they learned? }}We sought to determine if there was a change in tutors' self-reported confidence in their knowledge of the topic from pretest to posttest, as well as their perceived ability to apply what they learned. We compared tutors' Likert scale survey results collected prior to pretest and after posttest. Prior to beginning the lesson, tutors reported an average confidence level of 3.44 (\textit{SD} = 1.09) on a scale ranging from 1 (\textit{not at all confident}) to 5 (\textit{extremely confident}). Following participation in the lesson, this confidence level increased to 4.51 (\textit{SD} = 0.56). Among 81 tutors, 35 completed the post-lesson survey of confidence level. A paired sample \textit{t}-test of those 35 tutors revealed a statistically significant difference between pretest and posttest confidence level (t(34) = 6.82, p < .001), indicating a reliable improvement in tutors' self-reported confidence in their knowledge attending to students experiencing potential inequities. Additionally, tutors were asked to rate their confidence in applying what they learned after completing the short lesson. The average self-reported score for this measure was 4.71 (\textit{SD} = 0.46), indicating a high level of perceived ability to apply the acquired knowledge from the scenario to their own real-life tutoring.  

\subsection{\textbf{RQ3: How effective are large language models in assessing tutor's actions in responding to students managing possible inequity?}} GPT-4-turbo and GPT-4o showcased proficiency in evaluating tutor's actions in responding to students managing inequity, with better performance on the open responses tasking tutors to \textit{predict} the best approach compared to the open-response questions tasking tutors to \textit{explain} the rationale behind their provided approach. Table 5 displays the absolute performance of GPT-4-turbo and GPT-4o for \textit{predict} and \textit{explain} question types employing both zero- and few-shot prompting methods. Accuracy measures the proportion of correct predictions out of all predictions, providing an overall sense of a model's performance. AUC (Area Under the ROC Curve), often used with imbalance datasets, assesses how well the model distinguishes between classes, while the F1 score balances precision and recall, offering a measure of the model's effectiveness in handling both false positives and false negatives Given the same LLM, few-shot prompting outperformed zero-shot prompting. However, F1 scores for all models performed well ranging from 0.79 to 0.92.   

\begin{table}[ht]
\caption{Absolute model performance for open response questions prompting the models to assess tutor's ability to predict and explain the best tutor response.}
\begin{tabular}{lllllll}
\hline
\multirow{2}{*}{Model}  & \multicolumn{3}{c}{Predict} & \multicolumn{3}{c}{Explain} \\ \cline{2-7} 
                        & Acc.    & AUC     & F1      & Acc.    & AUC     & F1      \\ \hline
\texttt{GPT-4o} (zero-shot)     & 0.87    & 0.81    & 0.91    & 0.71    & 0.68    & 0.79    \\
\texttt{GPT-4o} (few-shot)      & 0.89    & 0.88    & 0.92    & 0.88    & 0.87    & 0.89    \\
\texttt{GPT-4-turbo} (zero-shot) & 0.85    & 0.90    & 0.79    & 0.80    & 0.78    & 0.83    \\
\texttt{GPT-4-turbo} (few-shot) & 0.89    & 0.87    & 0.92    & 0.89    & 0.89    & 0.90 \\\hline
\end{tabular}

\end{table}

\begin{figure*}[ht]
\includegraphics[width=\textwidth]{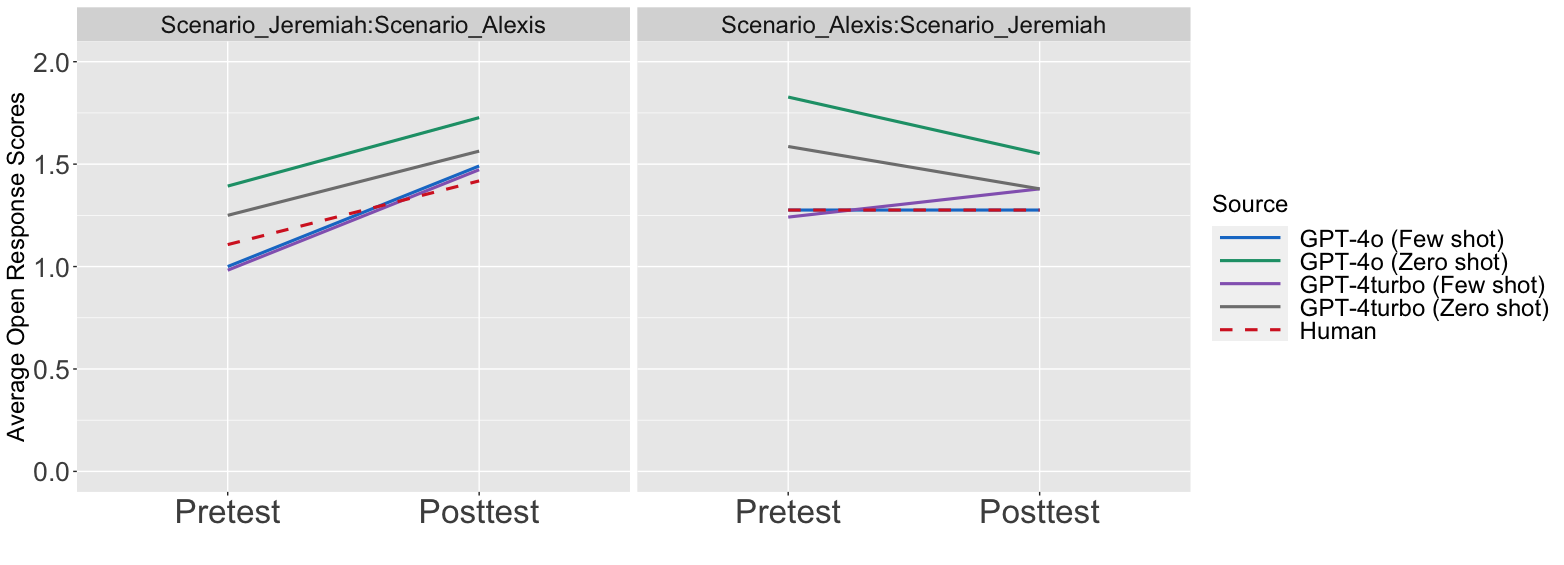}
\caption{Average open response scores (2 pts total) at pretest and posttest by scenario for each LLM model and human graders.} \label{fig:13}
\Description{Average open response scores (2 pts total) at pretest and posttest by scenario for each LLM model and human graders.} 
\end{figure*}

Figure \ref{fig:13} illustrates the average open responses scores (2 pts total) at pretest and posttest by scenario for each of the LLM models compared to human graders. Aligning with performance measures from Table 5, few-shot learning models more closely aligned with human graders compared to zero-shot learning models.  

\subsection{\textbf{RQ4: How do the large language models GPT-4o and GPT-4-turbo compare in performance, efficiency, and cost?}} Both GPT-4o and GPT-4-turbo performed well. For educational purposes at scale, it becomes necessary to determine the practical balance of performance, efficiency, and cost. For it doesn't matter how well a model performs if it is not feasible due to practicality and cost. Table 6 provides a comparison of model performance, efficiency, and estimated cost for the assessment of 1,000 tutor completions. Input and output token length includes pretest and posttest for the assessment of all open responses.\footnote{Input and output cost per token for GPT-4o (\$5/1M and \$15/1M) and GPT-4-turbo (\$10/1M and \$30/1M) as of Sept. 1, 2024 (https://openai.com/)}

\begin{table*}[ht]
\caption{Comparison of estimated cost for 1,000 lesson completions (includes pretest and posttest for all open-ended questions)}
\begin{tabular}{llllllll}
\hline
\textbf{Model}          & \textbf{input tokens} & \multicolumn{2}{l}{\textbf{input cost*}} & \textbf{avg. output tokens} & \multicolumn{2}{l}{\textbf{output cost*}} & \textbf{total cost} \\ \hline
\texttt{GPT-4o} (zero-shot)      & 696                   & \multicolumn{2}{l}{\$3.48}               & 188                         & \multicolumn{2}{l}{\$2.82}                & \$6.30                  \\
\texttt{GPT-4o} (few-shot)       & 1176                  & \multicolumn{2}{l}{\$5.88}               & 198                         & \multicolumn{2}{l}{\$2.97}                & \$8.85                  \\
\texttt{GPT-4-turbo} (zero-shot) & 696                   & \multicolumn{2}{l}{\$6.96}               & 271                         & \multicolumn{2}{l}{\$8.15}                & \$15.11                 \\
\texttt{GPT-4-turbo} (few-shot)  & 1176                  & \multicolumn{2}{l}{\$11.76}              & 282                         & \multicolumn{2}{l}{\$8.46}                & \$20.36 \\ \hline              
\end{tabular}

\noindent{*OpenAI API pricing as of Sept. 1, 2024 (https://openai.com/)
}
\end{table*}

\section{Discussion}
This study investigated the efficacy of various instructional conditions on tutor learning, the impact of assessment methods on learning outcomes, and the role of generative AI in evaluating tutor performance. Several important insights emerged, which offer a comprehensive understanding of this present work.
\subsection{\textbf{Tutors demonstrate new learning on applying equity-focused skills, however, improved balance of scenario difficulty is needed.}}
Tutors displayed evidence of new learning when responding to students possibly experiencing inequities from pretest to posttest. However, the practice related to learning gains in both item order conditions included the same activities, hence differences in learning gains by scenario order condition can be attributed to differences in students or the assessments of both scenario order conditions. One hypothesis for why learning gains were significantly higher in one scenario-order condition is that one of the scenario batteries is systematically different from the other. Tutors who had the Jeremy scenario followed by the Alexis scenario demonstrated a 12\% increase from the pretest to the posttest, respectively. This is not as high as previous work \cite{thomas2023tutor}, however, equity training is a more nuanced and ill-defined domain. The Alexis test battery had a higher average student score at pretest than the Jeremiah test batter and was therefore easier (79. 9\% compared to 72. 7\%), although not significantly so based on a two-sided \textit{t}-test, $t(12.87) = 0.88, p = .393$. Still, we tested the robustness of the main effect of time (i.e., learning gains) using an adjusted test score by applying z-score transformations within test-battery groups at pretest. That analysis consolidated a marginally significant learning gain between both scenario order conditions $F(1, 79) = 2.82, p = .097$. The interaction between scenario order condition and time remained robust after adjustment, $F(1, 79) = 3.96, p = .050$. Similarly, a Rasch model adjusting for individual item-difficulty and a fitted effect of time to test learning gains resulted in a marginally significant time effect representing learning gains, $\beta = 0.42, p = .066$.
\subsection{\textbf{Tutors reported improved confidence from pretest to posttest on applying equity-focused skills and are generally confident in applying what they have learned.}}The results indicate a significant improvement in tutors' self-reported confidence from pretest to posttest, highlighting the effectiveness of the lesson in increasing their confidence of attending to students experiencing potential inequities. The statistically significant increase in confidence suggests that the lesson had a meaningful impact on tutors' perceptions of their own competence. Furthermore, the high average confidence in applying the acquired knowledge (4.71) suggests that tutors feel well prepared to transfer what they learned into practice. However, does the increase in self-perceived confidence in applying learned skills translate into actual learning gains from pretest to posttest? To explore the relationship between tutor self-reported confidence in applying the skills taught in the lesson and their individual learning gains, we found no statistically significant relationship between the two. One possibility is that our statistical tests were underpowered given the post-survey dropout, and the true effect of the lessons on learning small.

\subsection{\textbf{Overall, Generative AI performed well, but not perfect, on assessing tutor performance.}}

The results show that GPT-4-turbo and GPT-4o are proficient at evaluating tutor responses, particularly in predicting the best approach when addressing student inequities. Few-shot prompting consistently outperformed zero-shot prompting across all metrics, demonstrating the value of providing examples to guide model responses \cite{brown2020language}. Strong F1 scores in all models (ranging from 0.79 to 0.92) suggest generally good performance. 

However, some notable challenges persist, particularly when evaluating responses that require subjective interpretation. Despite iteratively modifying the prompts, the models occasionally scored tutor responses differently from human graders. For instance, the following explanation was scored as incorrect (0) by human graders but coded as correct (1) by all models: \textit{``It'll help Jeremiah learn to take agency over his life.''} This response raises the question: Does the tutor's statement show that they recognize Jeremiah's need for support and encouragement in advocating for himself? According to GPT-4o's rationale, \textit{``The tutor's response indicates that they recognize the importance of Jeremiah learning to take agency over his life, which implies encouraging him to advocate for himself.''} This divergence in scoring highlights a fundamental difficulty in grading subjective responses. While the model's rationale is logical, it differs from the human graders' interpretation, potentially because human graders may consider broader contextual factors or expect more explicit language about advocacy and support. Alternatively, it's possible that the human grader's reasoning could be faulty, or influenced by biases or errors in judgment, which can happen in subjective evaluation scenarios. Determining the ``source of truth'' becomes especially challenging in such cases, as subjective responses often involve nuanced human judgment that LLMs, even with advanced prompting techniques, may struggle to fully capture \cite{condor2021automatic}. Moreover, human error can introduce inconsistencies, leading to disagreements between evaluators or between human and AI assessments. Therefore, developing more objective and transparent methods for determining the ``source of truth'' is advantageous.
\subsection{\textbf{Balancing performance, cost, and speed, GPT-4o with few-shot learning is the optimal choice.}}
\begin{figure*}[ht]
\includegraphics[width=\textwidth]{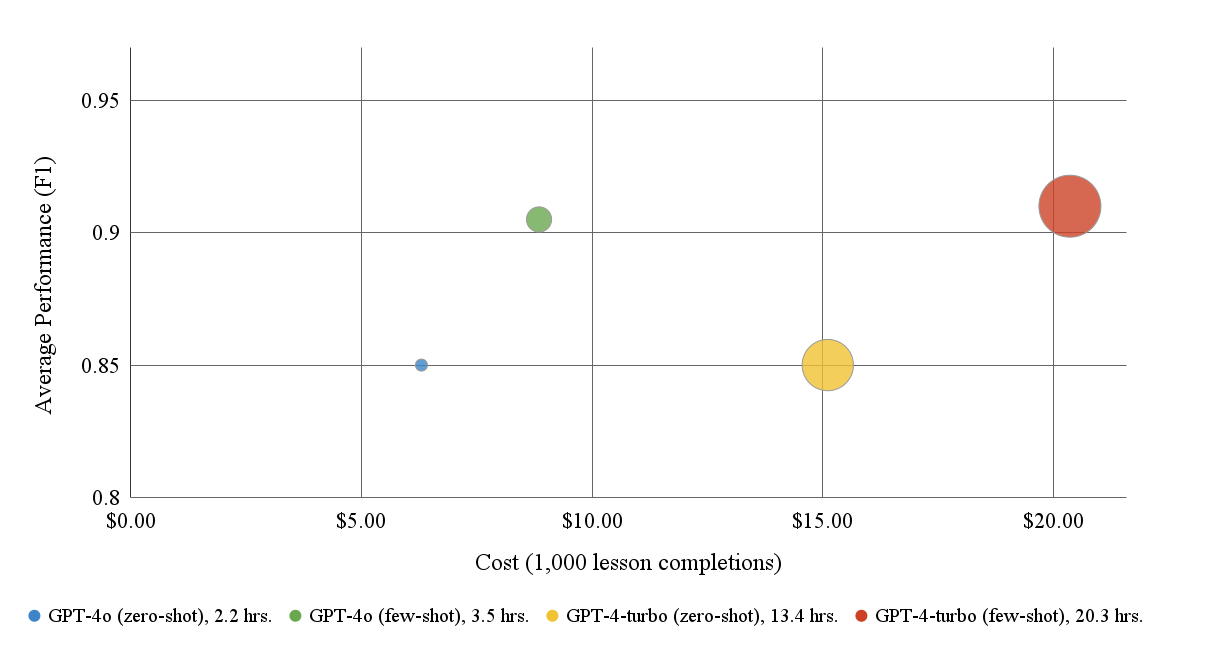}

\caption{Average performance and cost of GPT models for 1,000 lesson completions. Bubble size is proportional to processing time.} \label{fig:1}
\Description{Average performance and cost of GPT models for 1,000 lesson completions. Bubble size is proportional to processing time.}
\end{figure*}

Currently, GPT-4o with few-shot learning emerges as the preferred model when considering performance, cost, and speed, especially compared to GPT-4-turbo. While GPT-4-turbo offers only marginal performance improvements, it is nearly five times slower (20 tokens/sec vs. 109 tokens/sec for GPT-4o) and significantly more expensive \cite{openai_pricing}. Slow processing time would make it difficult to implement its use in providing automated assessment and immediate feedback to learners. Figure \ref{fig:1} displays the performance, cost, and processing time for each of the four models among 1,000 lesson completions. Using processing speeds and multiplying by the number of tokens (for simplicity combining input tokens and the average output tokens) for 1,000 lesson completions, we calculated the time it takes the models to assess tutor performance. GPT-4o using few-shot learning (represented by the green bubble in Figure \ref{fig:1}) exhibits the most favorable balance of performance, cost, and processing speed. In comparison, GPT-4-turbo models are substantially more expensive and slower. Using GPT-4o (few-shot), assessing the performance of 1,000 tutors completing the lesson would cost \$8.85 and take 3.5 hours. In contrast, What would it cost for humans to perform this same task? From our experience, human graders require around 15 seconds per response (1 minute per lesson for 4 open responses), resulting in 16.7 hours to assess 1,000 lessons. At a rate of \$30 per hour for a skilled human grader, the cost would be \$500. This comparison underscores the growing interest within learning analytics and the LAK community in leveraging generative AI for scalable, cost-effective grading and assessment. 

\section{Limitations}
This study has several limitations that merit consideration. First, the sample size of 81 tutor participants, while providing some insights, remains relatively small, potentially influencing the observed differences in learning gains across item counterbalancing conditions. Although we ruled out item difficulty as a primary factor, the small sample size raises the possibility that randomization differences between students could contribute to the gain differences. Future process-to-gain analyses could provide further clarity on this issue. Additionally, the reliability of the assessment used was lower compared to larger test batteries, potentially limiting the ability to detect learning gains with high precision. Despite this, we did observe marginally significant learning outcomes based on the available data. A notable challenge was the disparity in difficulty between the scenarios used in the pretest and posttest; while we applied an easiness correction to address this, creating scenarios that are analogous in learning objectives yet different enough to capture genuine learning gains remains a complex task. Regarding self-reported tutor data, only 35 out of 81 tutors completed the post-lesson survey, which presents a limitation to the generalizability of the findings for RQ2. The relatively low response rate may introduce bias, as the results could reflect the experiences of a subset of tutors who were more or less engaged or had a more positive or negative experience with the lesson. Another limitation stems from the nuanced nature of the scenarios, making human coding difficult, particularly in the alignment between human and AI evaluations for explain responses. Despite the challenges of human coding, the inter-rater reliability was high. There were possible ceiling effects, particularly among the selected-response questions. Using high-frequency, learner-sourced responses that were deemed ``incorrect,'' could be used as multiple-choice options \cite{thomas2023tutor}. Using high frequency and incorrect open responses as multiple-choice options may greatly increase lesson difficulty by capturing common misconceptions.

Only 35 out of 81 tutors completed the post-lesson survey, which presents a limitation to the generalizability of the findings for RQ2. The relatively low response rate may introduce bias, as the results could reflect the experiences of a subset of tutors who were more or less engaged or had a more positive or negative experience with the lesson. Another limitation stems from the nuanced nature of the scenarios, making human coding difficult, particularly in the alignment between human and AI evaluations for \textit{explain} responses. Despite the challenges of human coding, the inter-rater reliability was high. There were possible ceiling effects, particularly among the selected-response questions. Using high-frequency, learner-sourced responses that were deemed ``incorrect,'' could be used as selected-response options \cite{thomas2023tutor}. Using common and incorrect open-ended responses as selected response options may greatly increase lesson difficulty by capturing common misconceptions.
\section{Future Work and Conclusion}
Future work should address several key areas. First, increasing the complexity of scenarios, particularly the Alexis scenario, could provide deeper insights into tutor performance. Furthermore, research could investigate how increased self-reported confidence translates into real-world tutoring effectiveness and long-term student outcomes. Further studies should also explore the performance and variability of all combinations of few-shot prompts, rather than only reporting the best iteration, to better understand the impact of prompt design. Another important direction is determining the predictive validity of lessons and tutor learning transfer by analyzing real-life tutor-student interactions and transcription data. Finally, exploring more objective metrics, such as multiple-choice questions, as a potential ``source of truth'' could enhance the robustness of assessment methods.

In conclusion, this study demonstrates the potential of generative AI, particularly GPT-4o with few-shot prompting, as a valuable tool to assess the performance of the tutor in equity-focused online lessons. Using a mixed methods approach, we evaluated the performance of 81 undergraduate remote tutors, combining quantitative analysis of learning gains and self-reported confidence with the assessment of tutor responses. Tutors showed marginally significant learning gains and reported increased confidence from pretest to posttest in applying skills to address inequitable situations, underscoring the effectiveness of the training. While generative AI models performed well in evaluating tutor responses, challenges remain, particularly in improving alignment between scenario difficulty and model assessment accuracy. By balancing cost, performance, and speed, GPT-4o emerged as the most effective model for large-scale assessment. Future work should focus on refining scenario complexity and optimizing LLM prompts to enhance tutor training and AI-driven assessments. The release of the data set from this study offers a valuable resource for further research on equity-focused learning and assessment.

\begin{acks}
 This work was made possible with the support of the Learning Engineering Virtual Institute. The opinions, findings and conclusions expressed in this material are those of the authors.
\end{acks}

\bibliographystyle{ACM-Reference-Format}
\bibliography{main}

\appendix

\section{Digital Appendix}
All analysis code, study materials, and log data references can be found in the study's supplementary GitHub repository:\\
\href{https://github.com/CMU-PLUS/LAK2025-Inequity}{https://github.com/CMU-PLUS/LAK2025-Inequity}

\end{document}